\documentclass[twocolumn,prl,aps,showpacs,footinbib,amsmath,amssymb,superscriptaddress,longbibliography]{revtex4-1}
\usepackage{graphicx}
\usepackage{dcolumn}
\usepackage{xcolor}
\usepackage{bm}
\usepackage{hyperref}
\usepackage{amsmath,amssymb,amsfonts,bbold}
\usepackage[normalem]{ulem}
\usepackage{miller}
\usepackage{todonotes}
\usepackage{physics}
\graphicspath{{figures/}}

\definecolor{dartmouthgreen}{rgb}{0.05, 0.5, 0.06}
\definecolor{darkspringgreen}{rgb}{0.09, 0.45, 0.27}
\definecolor{DebianRed}{rgb}{0.84, 0.04, 0.33}
\definecolor{darkpowderblue}{rgb}{0.0, 0.2, 0.6}
\definecolor{lightgreen}{rgb}{0.0, 1.0, 0.0}

\begin{document}

\preprint{}

\title{Coupling of Yu-Shiba-Rusinov states in 1D chains of Fe atoms on Nb(110)}

\author{Felix Friedrich}
\email[]{felix.friedrich@physik.uni-wuerzburg.de}
\author{Robin Boshuis}
\affiliation{Physikalisches Institut, Experimentelle Physik II,
Universit\"at W\"urzburg, Am Hubland, 97074 W\"urzburg, Germany}
\author{Matthias Bode}
\affiliation{Physikalisches Institut, Experimentelle Physik II,
Universit\"at W\"urzburg, Am Hubland, 97074 W\"urzburg, Germany}
\affiliation{Wilhelm Conrad R\"ontgen-Center for Complex Material Systems (RCCM), Universit\"at W\"urzburg, Am Hubland, 97074 W\"urzburg, Germany}
\author{Artem Odobesko}
\affiliation{Physikalisches Institut, Experimentelle Physik II,
Universit\"at W\"urzburg, Am Hubland, 97074 W\"urzburg, Germany}

\date{\today}

\begin{abstract}
The hybridization of Yu-Shiba-Rusinov states in dimers of magnetic impurities leads to molecular-like bonding and antibonding modes. In many-impurity systems, the interaction gives rise to YSR bands and can even result in the formation of a topologically non-trivial superconducting state, characterized by Majorana fermions at the edges of the system. To obtain a more detailed understanding of these interactions, we investigate the coupling of YSR
states in short one-dimensional Fe chains on clean Nb(110). 
We observe a splitting of the single-atom YSR peaks into multiple states 
with even or odd spatial symmetry and identify a peculiar dependence 
of the even and odd states' energy position on the chain length.
\end{abstract}


\maketitle

\section{Introduction}
While the presence of the later so-called Yu-Shiba-Rusinov (YSR) states 
was already theoretically predicted in the 1960s~\cite{yu1965,shiba1968,rusinov1969}, 
their first direct experimental observation was only accomplished 30 years later~\cite{yazdani1997}. 
In the recent past, these bound states, which appear at the site of magnetic impurities 
in a superconducting host material, gained renewed interest in the field of topological superconductivity. 
The proposal  to realize Majorana fermions (MFs) in a one-dimensional solid-state system with helical spin structure
by using chains of magnetic adatoms on an s-wave superconductor \cite{nadj-perge2013,braunecker2013,pientka2013,klinovaja2013} triggered a surge of experimental and theoretical contributions
to this field~\cite{nadj-perge2014,peng2015,ruby2017,feldman2017,jeon2017,kim2018}.

In addition to MFs, which are bound to the Fermi energy and localized at the chain ends, 
the hybridization of YSR states leads to the formation of extended YSR bands~\cite{ruby2015,schneider2021}. 
This coupling between YSR states has extensively been studied for dimers of magnetic impurities in theoretical and experimental works: Single YSR states of individual atoms were predicted to split into two states with wave functions of even and odd spatial symmetry, 
similar to the formation of bonds in the H$_2^+$-molecule~\cite{flatte2000}. 
Depending on the atom--substrate interaction and the distance of the two atoms relative to the Fermi wavelength, 
either the even or odd state is lower in energy and the dimer can undergo a series of phase transitions changing this order~\cite{morr2003}. 
Experimentally, the splitting of YSR states was demonstrated, e.g., in Mn dimers on Pb(111)~\cite{ji2008} 
and cobalt phthalocyanine dimers on NbSe$_2$~\cite{kezilebieke2018}. 
The splitting of YSR states associated with individual orbitals was observed in Mn dimers on Pb(001)~\cite{ruby2018}. 
With the help of density functional theory (DFT) calculations, Choi \textit{et al.}~\cite{choi2018} were able to show 
that a splitting of YSR states in Cr on \mbox{$\beta$-Bi$_2$Pd} only occurs for ferromagnetically (FM) coupled dimers, 
whereas antiferromagnetically (AFM) coupled dimers did not show any splitting, 
as theoretically predicted~\cite{flatte2000,morr2003}. 
However, recent experiments suggest that spin-orbit coupling in the substrate, 
which was not considered in previous theoretical calculations, also allows for a splitting in AFM coupled dimers~\cite{beck2020}.

In this contribution, we extend the investigation of the coupling 
of YSR states to one-dimensional chains consisting of up to four atoms. 
We first study the coupling of YSR states for Fe atom dimers on clean Nb(110) with various inter-atomic spacings and dimer orientations. 
STM tip functionalization with single CO molecules allows us 
to topographically resolve even dense-packed dimer configurations on the Nb substrate, 
which show clear splitting of the single-atom YSR peak into an even and an odd state. 
With increasing chain length, the splitting of the YSR states becomes more complex. 
Spatially resolved maps of the differential conductance indicate that the symmetry 
of the energetically lowest and highest state alternates between chains with an odd or even number of atoms. 
The results are rationalized by a simple quantum-mechanical model.

\section{Experimental methods}
All measurements are performed in a home-built low temperature STM at a base temperature of 1.4\,K. 
The Nb(110) crystal is cleaned by sputtering with Ar ions 
and a series of high temperature flashes~\cite{odobesko2019}. 
The Fe atoms are deposited in-situ onto the cold Nb substrate at a temperature of 4.2\,K. During the deposition, small clusters including one-dimensional chains of up to four atoms self-assemble on the clean surface.

In order to increase the spatial resolution of the STM experiments, 
we functionalize electro-chemically etched tungsten tips with a CO molecule on the tip apex. 
To pick up a CO molecule from a clean Cu(001) surface, we stabilize the tip 
at $U=-2\,\mathrm{V}$ and $I=1\,\mathrm{nA}$ above the molecule, switch off the feedback loop 
and change the bias voltage to $-4\,\mathrm{V}$, similar to the procedure described in Ref.~\onlinecite{bartels1998}. The successful transfer of the CO molecule to the tip apex is indicated 
by a sudden jump of the tunneling current to a lower absolute value.
To verify that the CO molecule is still on the tip apex after exchanging samples and approaching the Nb(110) surface, 
we measure the second derivative of the tunneling current on top of clean Nb(110). 
Only if the CO molecule is still present at the tip apex the excitation of the hindered rotation and hindered translation mode via inelastic tunneling 
leads to characteristic, electron--hole-symmetric peaks in the $\mathrm{d}^2I/\mathrm{d}U^2$ signal~\cite{lauhon1999,heinrich2002,supplmat}.
After recording the topographic information of the Nb(110) surface 
and the self-assembled Fe adatom clusters, we remove the CO atom from the tip apex and gently dip the tip into the surface.
The resulting material transfer from the Nb sample onto the tip apex renders the tip superconducting, 
thereby increasing the spectroscopic resolution~\cite{pan1998,franke2011}.
Since the presence of a superconducting gap~$\Delta_\mathrm{tip}$ in the spectral function of the tip 
causes a corresponding shift of features in the sample's local density of states (LDOS) in the measured spectra, 
a precise knowledge of the gap size is required for the later interpretation of the data.  
The value of $\Delta_\mathrm{tip}$ is obtained by fitting the spectrum measured on the clean surface 
and is indicated in the figure captions (see Ref.~\onlinecite{odobesko2020} for the fitting procedure). 
All spectroscopic measurements are performed with a modulation voltage of 0.1\,mV at a modulation frequency of 890\,Hz.

\section{Results}
\begin{figure}
\includegraphics[scale=1]{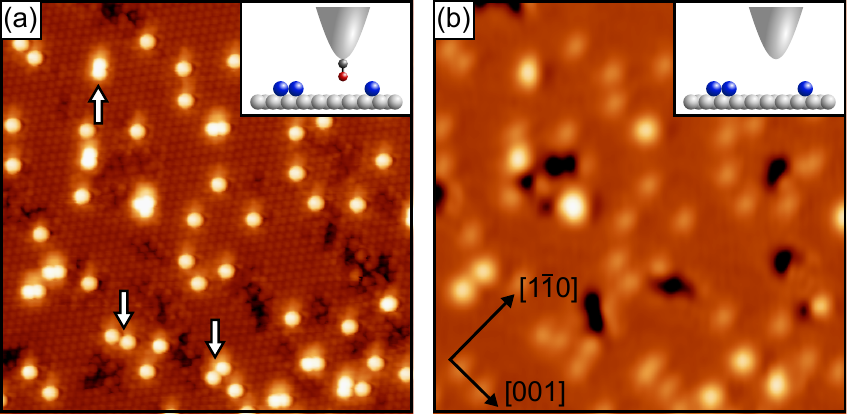}
\caption{Topographic images of the same Nb(110) surface area (scan range: 15$\times$15\,nm$^2$) 
		recorded with (a) the CO-functionalized and (b) the bare metal tip. 
		The CO-functionalized tip uncovers the Nb lattice and the individual Fe atoms in the self-assembled clusters.  
		Scan parameters: \mbox{(a) $U = -10\,\mathrm{mV}$,} \mbox{$I = 5\,\mathrm{nA}$;} 
		\mbox{(b) $U = -1\,\mathrm{V}$,} \mbox{$I = 400\,\mathrm{pA}$}.}
\label{fig:surface}
\end{figure}

\begin{figure}
\includegraphics[scale=1]{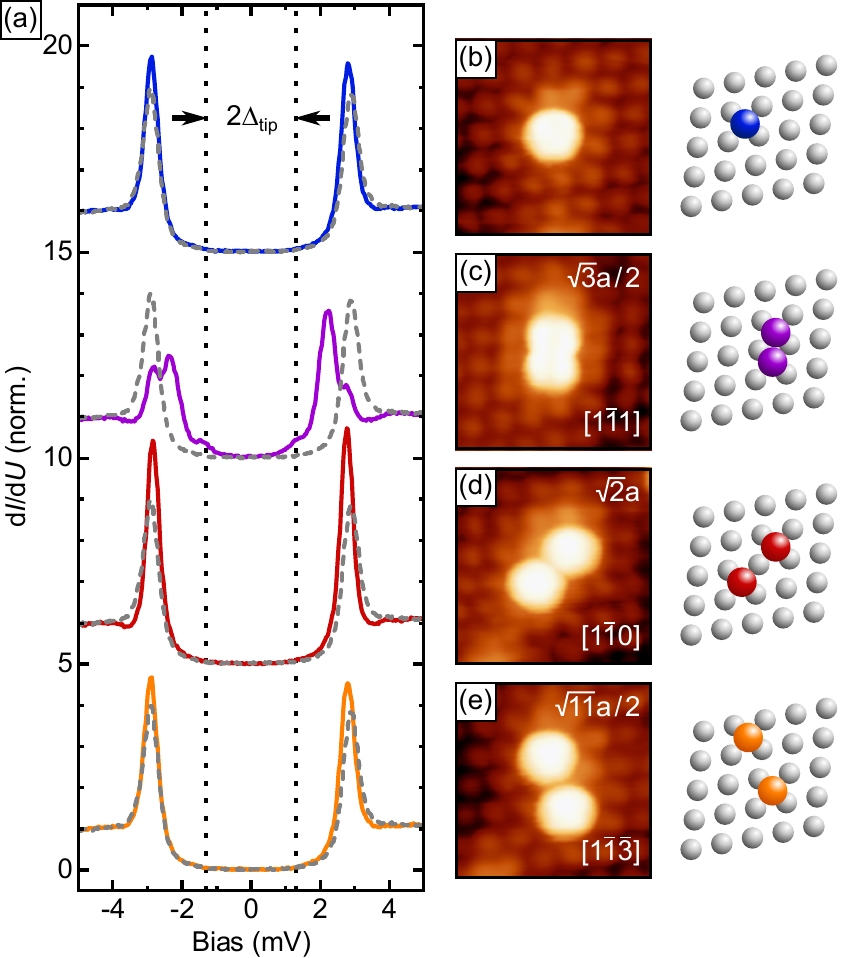}
\caption{(a) Tunneling spectra of a single Fe atom and the three different dimers with the shortest inter-atomic distance. 
		The gray hatched line represents the spectrum on clean Nb(110).  Spectra are vertically shifted for clarity. 
		The gap width of the superconducting tip amounts to \mbox{$\Delta_\mathrm{tip} = 1.31\,\mathrm{meV}$} and is indicated by dotted lines.  
		\mbox{(b)-(e)} High-resolution STM images (scan range:  $2\times 2\,\mathrm{nm}^2$)
		of the respective dimers (left) and their ball model representation (right). 
		In each panel the dimer length and orientation is indicated.
		Stabilization parameters: \mbox{(a) $U = -7\,\mathrm{mV}$,} \mbox{$I = 400\,\mathrm{pA}$;} 
		\mbox{(b)-(e) $U = -10\,\mathrm{mV}$,} \mbox{$I = 5\,\mathrm{nA}$.}  
\label{fig:spectra}}
\end{figure}

Figure~\ref{fig:surface} shows a constant-current STM image of the clean Nb(110) surface with deposited Fe atoms. 
The data in Fig.~\ref{fig:surface}(a) recorded with a CO-functionalized tip 
clearly reveals the Nb(110) lattice with the lattice constant of $a = 3.3\,\mathrm{\AA}$. 
Oxygen and hydrogen impurities appear as dark defects on the surface. 
Fe atoms and small Fe clusters show up as bright protrusions and can clearly be resolved.
We find dimers oriented along three different crystallographic directions of the Nb(110) surface (excluding double counting due to mirror symmetries), namely the \hkl[1-11], \hkl[1-10], 
and the \hkl[1-13] direction, which are marked by arrows. 
Close-ups of representative dimers are shown in Fig.~\ref{fig:spectra}\mbox{(b)-(e)}.  
Surprisingly, we do not observe any self-assembled \hkl[001]-oriented dimers on the surface~\cite{atomManipulation}.
Fig.~\ref{fig:surface}(b) shows an image of the same area recorded after removing the CO molecule from the tip. 
The Fe atoms appear less sharp and individual atoms within clusters cannot be resolved anymore. 
As the scan is recorded at a bias voltage of $-1\,\mathrm{V}$, hydrogen is not visible 
and the remaining dark spots indicate the location of oxygen on the Nb surface~\cite{odobesko2019}. 
All spectroscopic measurements shown below are performed on atoms and clusters not sitting on top of oxygen-reconstructed Nb, as the presence of oxygen changes the coupling of Fe atoms to the Nb substrate 
and hence influences the energy of the YSR bound states of these atoms~\cite{odobesko2020,oxygenInfluence}. 
Hydrogen in contrast has no detectable influence on the superconducting LDOS.

Representative spectra measured on a single Fe atom and on top
of the various Fe dimers are displayed in Fig.~\ref{fig:spectra}(a). 
For comparison, the spectrum measured on clean Nb(110) is depicted as a gray dashed line. 
In Fig.~\ref{fig:spectra}\mbox{(b)-(e)} scans of the respective atom/dimer are presented 
together with a ball model that demonstrates the dimer orientations on the substrate. 
The actual adsorption sites of the single atoms on the lattice might differ 
from the four-fold hollow site predicted by DFT~\cite{odobesko2020}, which is displayed 
in the ball model representation, see right panels of Fig.~\ref{fig:spectra}\mbox{(b)-(e)}. 
This is especially true for dimers with a short inter-atomic distance, 
as interactions can lead to a substantial inwards displacement of the atoms~\cite{bode1996}. 
Indeed, our high spatial resolution measurements suggest small shifts 
of the atoms' charge density in nearest-neighbor dimers oriented along the \hkl[1 -1 1] direction~\cite{supplmat}.

As reported previously, close to the Fermi level the LDOS of a single Fe atom on clean Nb(110) 
only slightly differs from that of the bare substrate~\cite{odobesko2020}. 
The small enhancement of the $\dd I/\dd U$ signal at the coherence peaks in comparison to the clean Nb(110) surface indicates the presence of YSR states 
close to the superconducting gap edge with an energy of $E_\mathrm{YSR} \approx \Delta_\mathrm{Nb}$. 
In contrast to this, the \hkl[1-11]-oriented dimer, which has the shortest inter-atomic distance 
[cf.\ Fig.~\ref{fig:spectra}(c)], shows two pairs of additional in-gap peaks. 
The more intense pair appears at a bias voltage of $2.3 \pm 0.1\,\mathrm{mV}$ in the convoluted tip and sample LDOS, 
the weaker one is detected at a bias voltage that almost perfectly corresponds to the tip gap, 
i.e., the peaks are energetically located close to the Fermi level in the sample LDOS. 
As will be discussed below, the peak intensity strongly depends on the tip's lateral position relative to the dimer. 

The LDOS of dimers with larger inter-atomic distances [cf.\ Fig.~\ref{fig:spectra}(d) and (e)] lacks any additional peaks. 
However, the spectrum of the dimer oriented along the \hkl[1 -1 0] direction, Fig.~\ref{fig:spectra}(d), 
still clearly differs from that of a single Fe atom, showing a much higher intensity at the position of the coherence peaks. 
Spectra measured on dimers with a distance equivalent to that of \hkl[1-1-3]-oriented dimers, Fig.~\ref{fig:spectra}(e), 
or larger (not shown here) strongly resemble those of single atoms.

\begin{figure*}
\includegraphics[scale=1]{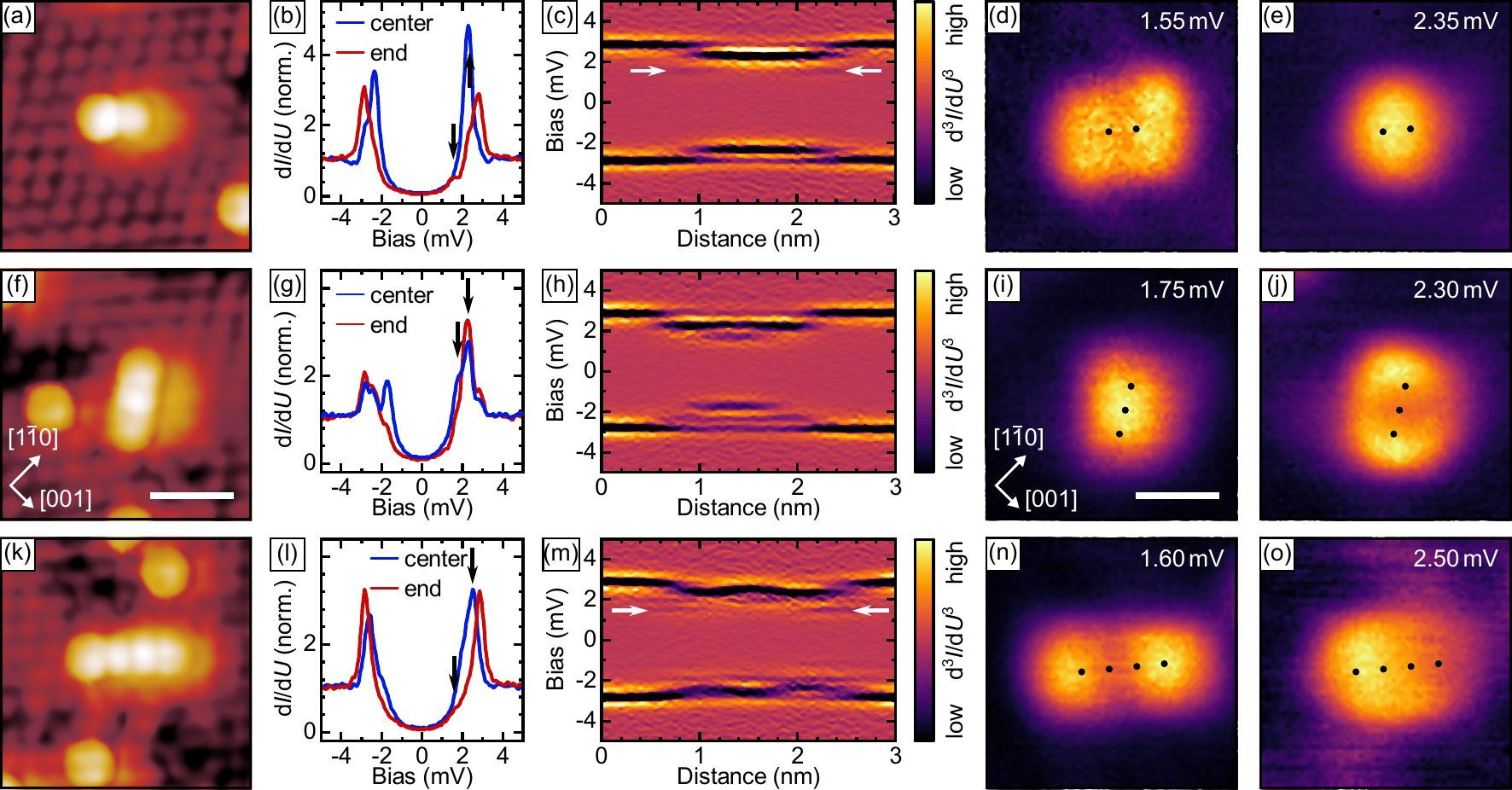}
\caption{\label{fig:fullgrid}(a) Topographic image of a dimer along the nearest-neighbor \hkl[1-11] direction. 
		(b) Spectra taken at the center and on the end of the dimer. 
		(c) Numerically calculated second derivative of the $\dd I/\dd U$ signal measured along the dimer.  
		White arrows highlight the low intensity states. 
		(d),(e) $\dd I/\dd U$ maps at indicated bias voltages corresponding to the peak positions marked by black arrows in (b). 
		Black dots symbolize the positions of the Fe atoms. \mbox{$\Delta_\mathrm{tip} = 1.33\,\mathrm{meV}$}. 
		Stabilization parameters: (a) \mbox{$U = -10\,\mathrm{mV}$,} \mbox{$I = 3\,\mathrm{nA}$}; 
		(b)-(e) \mbox{$U = -7\,\mathrm{mV}$,} \mbox{$I = 1\,\mathrm{nA}$.} (f)-(j) 
		and (k)-(o) are same as (a)-(e), but for a trimer and tetramer, respectively. 
		The surface orientation of all scans is given in (f) and (i). Scale bars are $1\,\mathrm{nm}$.}
\end{figure*}

To experimentally assess the splitting of hybridized YSR states and their wave functions in short one-dimensional chains we perform full spectroscopy measurements on chains of two, three, and four atoms along the \hkl<1 -1 1> direction. Tunneling spectra (voltage range $\pm 7\,\mathrm{mV}$) 
are recorded every $0.5\,\mathrm{\AA}$ on a $3 \times 3\,\mathrm{nm}^2$ scan range.
In Fig.~\ref{fig:fullgrid} we present the spatial distribution of YSR states in a linear \mbox{(a)-(e)} dimer, 
\mbox{(f)-(j)} trimer and \mbox{(k)-(o)} tetramer oriented along the nearest-neighbor \hkl<1-11> direction. 
The distance of the investigated chains to other Fe atoms is significantly larger 
than the maximum coupling distance determined from the measurements on dimers. 
Note that the CO-functionalized tip utilized in this set of measurements produces a subtle shadow 
on the right side of all atoms as a result of a slight CO/metal double tip. 
We would like to emphasize that the data presented in the following are not influenced by this artifacts, 
since---as mentioned before---the tip was re-conditioned on the Nb sample prior to the spectroscopic measurements~\cite{supplmat}.

As discussed, the Fe dimer exhibits two pairs of YSR in-gap peaks with a higher intensity of the electron-like states that are clearly visible at positive bias in Fig.~\ref{fig:fullgrid}(b). 
In Fig.~\ref{fig:fullgrid}(c) we show the second numerical derivative of the $\dd I/\dd U$ signal measured along the dimer. 
Peaks in the $\dd I/\dd U$ signal appear as dark areas in the color-coded second derivative. 
As indicated by white arrows in the graph, the low-energy YSR state appears 
at the ends of the dimer and shows a vanishing intensity in between. 
The high-energy state is localized at the dimer center. 
The $\dd I/\dd U$ maps presented in Fig.~\ref{fig:fullgrid}(d) and (e) which are measured 
at bias voltages~$U = (\Delta_\mathrm{tip} + E_\mathrm{YSR})/\mathrm{e}$ emphasize this observation. 
We can hence assign an even (odd) symmetry to the high- (low-)energy YSR state of the Fe dimer.

The spectra measured on the trimer displayed in Fig.~\ref{fig:fullgrid}(f) 
also exhibit two pairs of in-gap states, as shown in Fig.~\ref{fig:fullgrid}(g) and (h). 
Both YSR states are well pronounced but separated by a smaller energy than those in the dimer. 
In contrast to the dimer, the low-energy state has an intensity maximum at the center of the chain, 
whereas the high-energy state is more pronounced at the chain ends. 
The $\dd I/\dd U$ map in panel~(i) (panel~(j)) of Fig.~\ref{fig:fullgrid} again demonstrates 
the non-split (split) intensity distribution of the low- (high-)energy YSR state, 
characteristic for the even (odd) state~\cite{morr2006}. 
Note that the apparent rotation of the high-energy state's intensity distribution away from the chain 
towards the high symmetry \hkl[001] direction is reproducible for all trimers along the \hkl[1-11] and \hkl[1-1-1] direction (not shown here).

In Fig.~\ref{fig:fullgrid}(k)-(o) we display the results obtained for a chain of four Fe atoms along the \hkl[1-11] direction. 
In panel (l), two spectra measured with the STM tip positioned at the center and the end of the chain are shown. 
In comparison to the peaks observed on the dimer and trimer, the peaks at $\pm 2.5\,\mathrm{mV}$ in the spectrum taken at the tetramer center are rather broad. 
As can be seen in panel (m), these peaks are not constant in energy along the chain 
but shift from a lower absolute energy at the chain ends to a higher energy at the chain center. 
We will discuss possible explanations for this observation in the following section. 
The $\dd I/\dd U$ map shown in Fig.~\ref{fig:fullgrid}(o) is recorded at a bias voltage 
which corresponds to the energy of the high-intensity peak at the chain center [cf.\ panel~(l), blue line]. 
It reveals a rather homogeneous $\dd I/\dd U$ intensity along the chain.
However, if the spectroscopic measurements are evaluated at a slightly lower bias voltage corresponding to the energetic position of the same peak further at the chain end, the intensity distribution shows a minimum
at the chain center [see Fig.~\ref{fig:fullgrid}(m) and Fig.~S1(d) in Ref.\,\onlinecite{supplmat}].
The low-energy state indicated by white arrows in Fig.~\ref{fig:fullgrid}(m) is localized 
at the end of the chain and exhibits a vanishing intensity at the center of the tetramer [see Fig.~\ref{fig:fullgrid}(n)].

We find that increasing the chain length from a dimer to a trimer reverses the energetic position of the even and odd state. 
For a tetramer, the situation is more complicated, but we can state that the YSR state with the highest energy 
is localized at the center of the chain, whereas the state with the lowest energy is localized at the chain ends. 
We consequently identify an oscillatory behavior 
of the energetically lowest/highest state's spatial symmetry as a function of chain length.

\section{Discussion}

As mentioned above, the splitting of single-atom YSR states into one odd and one even state in dimers 
has been studied in detail theoretically~\cite{flatte2000,morr2003,yao2014,meng2015} and experimentally~\cite{ji2008,kezilebieke2018,ruby2018,choi2018,beck2020}.
For the coupling of three magnetic atoms one expects a splitting of the single-atom resonance into three separate states, 
two of which have an even and one an odd symmetry~\cite{morr2006}. 
The tight-binding model employed by Morr and Yoon~\cite{morr2006} to calculate 
the energies and symmetries of YSR states for dimers and trimers can easily be extended to longer chains. 
The approach is based on the linear combination of atomic orbitals (LCAO) and for an $n$-atomic chain 
only contains the single-atom YSR energy~$E_0$ and the $n-1$ hopping constants $k_\mathrm{i}$ as free parameters. 

As we can deduct from the experimental data on Fe dimers on clean Nb(110) (cf.\ Fig.~\ref{fig:spectra}), 
the coupling between atoms with a distance of $\sqrt{11}a/2 = 5.5\,\mathrm{\AA}$ or larger is negligable. 
This is due to (i) the weak interaction of the single Fe atoms with the itinerant quasiparticles~\cite{odobesko2020}, which mediate the coupling between the adatoms,
and (ii) the small overlap of the single-atom YSR wave functions, resulting from their small spatial extent.
The latter is limited by the dimensionality of the involved bands~\cite{menard2015,kim2020}, 
which have 3D character for the Nb(110) surface~\cite{odobesko2019}, and the Fermi wavelength~$\lambda_\mathrm{F}$, causing a decay of the wave function proportional to $\lambda_\mathrm{F}/r$~\cite{rusinov1969}.  
Since the Fermi surface of Nb(110) contains an extensive number of bands~\cite{odobesko2020_vortices} 
that might be involved in the coupling of YSR states,
we cannot determine $\lambda_\mathrm{F}$ for our system. 
However, our measurements indicate that the YSR states are strictly localized to the location of the Fe atom and are not detectable on the bare substrate surrounding it (see Fig.\ S1(a) in~\cite{supplmat}).
 
We thus do not expect coupling between next-nearest neighbor atoms in the chain, 
which have an inter-atomic distance of $2\cdot\sqrt{3}a/2=5.7\,\mathrm{\AA}$, 
i.e., even larger than the inter-atomic distance of $\sqrt{11}a/2 = 5.5\,\mathrm{\AA}$ over which no coupling is observed (cf.\ Fig.~\ref{fig:spectra}).
In our model we can, therefore, set the hopping terms between all but the neighboring atoms, i.e. $k_2$ and $k_3$, to zero.

\begin{figure}
\includegraphics[scale=1]{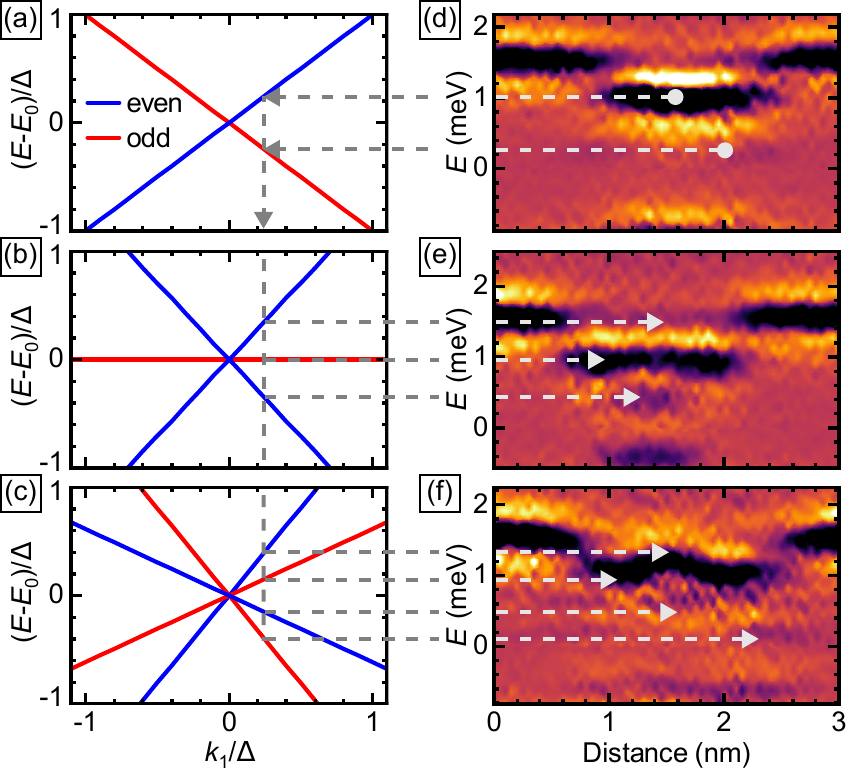}
\caption{(a)-(c) Calculated energy splitting for electron-like states as a function of the nearest-neighbor hopping term~$k_1$ for a dimer, trimer and tetramer, respectively. Hopping between next-nearest neighbor atoms is neglected. 
			The symmetry is reflected by the color of the lines with blue indicating the even and red the odd states. 
			(d)-(f) Energy of YSR states at positive bias extracted from the data in Fig.~\ref{fig:fullgrid} (c), (h) and (m) by subtracting $\Delta_\mathrm{tip}$. The data is shifted to align the center energy of the experimental results with ``0'' in the calculations (see main text for details).  Length of the $y$-axis is $2\Delta_\mathrm{Nb}$, the same as in (a)-(c). The dimer is used as a reference to fix $k_1$ in the calculations, such that calculations and experiment match perfectly for the dimer. As indicated by the arrows, the calculation results at the extracted value of $k_1$ reproduce the experimental results for the trimer and tetramer.}
\label{fig:lcoaModel}
\end{figure}

In Fig.~\ref{fig:lcoaModel} (a)-(c) we present the calculated shift of the YSR energies from $E_0$ for the electron-like states 
in a dimer, trimer and tetramer as a function of the nearest-neighbor hopping term~$k_1$. 
If only nearest-neighbor hopping is considered as discussed above, the states split symmetrically around $E_0$. 
As expected, the number of states is equal to the number of atoms in the chain. 
Further, we find that the spatial symmetry of the states alters between even and odd in energy, 
where the state highest in energy is always even for positive $k_1$. 

To map the model to our experimental results we first need to discuss the behavior of $E_0$ in our measurements. 
In contrast to the model, the center energy, i.e. the energy at the center of the energetically highest and lowest YSR state, 
is not equal to the single-atom YSR state energy in our experiments. 
This is expected from more elaborate calculations which include shifts originating from Coulomb-like interactions and the overlap of the wave functions~\cite{ruby2018}. Further shifting of the center energy can be caused by the relative alignment and coupling of the atom spins~\cite{morr2003} or a modification of the atomic adsorption site within the chains, leading to a different interaction with the substrate.
In order to account for these effects, we extract the center energy~$E_0'$ from our measurements in order to align our experimental results with the calculations.
We find $E_0' = (0.6 \pm 0.1)\,\mathrm{meV}$ for the dimer, $E_0' = (1.0 \pm 0.1)\,\mathrm{meV}$ for the trimer 
(note that the center energy is equal to the energy of the odd YSR state according to the model) 
and $E_0' = (0.7 \pm 0.1)\,\mathrm{meV}$ for the tetramer.

We display zooms of the $\dd^3I/\dd U^3$ line grids presented in Fig.~\ref{fig:fullgrid}(c), (h), and (m) 
next to the model calculations in Fig.~\ref{fig:lcoaModel}. 
For the dimer the tight-binding model correctly reproduces both the even and the odd state observed in our measurements. 
We can thus use the experimental data to fix the hopping term $k_1$ in our calculations, 
which, in the simplest scenario, is expected to be the same for all chains. Since for the dimer $E^\mathrm{even} - E^\mathrm{odd}$ is equal to $2k_1$, we obtain $k_1 \approx (0.40\pm0.04)\,\mathrm{meV}\approx(0.26\pm0.03)\Delta_\mathrm{Nb}$~\cite{hoppingTerm}.  
We can now extract the energy splitting of the YSR states for the trimer and tetramer from the calculations and compare them to our experimental data. 

The derived values are indicated by arrows in Fig.~\ref{fig:lcoaModel}(b,e) and (c,f) for the trimer and tetramer, respectively.
For the trimer we find both good qualitative and quantitative agreement between the experimental data and the model. 
The experimentally resolved odd and even state are well represented by our calculations. 
Further, an even state close to the coherence peaks is predicted, which we do not observe in our measurements. 
In general, a single pair of in-gap states sitting close to the coherence peaks is expected to increase the measured tunneling conductance at this energy, similar to our findings on single Fe atoms. If the states are located well inside the gap, however, the LDOS at the gap edge is decreased in favor of the in-gap states. We can thus, due to the additional in-gap states of the trimer, not disentangle coherence peaks and possible YSR states at the same energy.

When the fourth atom is added to the chain, the number of states in the model also increases to four. 
However, we only observe two clear peaks inside the gap. 
The low-energy state at $\approx 0.2\,\mathrm{meV}$ has an odd symmetry, in agreement with our calculations. 
As mentioned in the previous section, the peak at higher energy is rather broad and appears to shift in energy along the chain. 
Considering the results of our calculations, we speculate that this peak consists of several peaks 
of different intensity that are smeared into a single peak due to thermal broadening. 
A closer look at the second derivative of the measured spectra in Fig.~\ref{fig:lcoaModel}(f) suggests 
that the wave-like structure around $\approx1.2\,\mathrm{meV}$ can be interpreted as the overlap of one even (energetically higher) and one odd (energetically lower) state. Additionally we identify a very weak fourth state at $\approx0.6\,\mathrm{meV}$. 
However, since $\dd I/\dd U$ maps at this energy are dominated by the high-intensity odd state 
at $\approx1.0\,\mathrm{meV}$, the symmetry of this fourth state is difficult to characterize. 

Although higher spectral resolution is required to confirm the presence and symmetry of all states in the tetramer, 
the coupling of YSR states in our one-dimensional system seems to  be well described by our calculations. 
Especially the observed symmetry reversal of the states with highest and lowest energy 
as a function of chain length is clearly reflected in our model. 
As the system becomes more complex, however, the quantitative agreement is limited by the simplicity of the model, 
which is evident from the too large splitting predicted for the tetramer [cf.\ Fig.~\ref{fig:lcoaModel}(f)]. 
Further, as shown by more elaborate calculations, the observed symmetry reversal is not universal 
and depends---among other parameters---on the ratio of the inter-atomic distance to the Fermi wavelength~\cite{supplmat}. 

\section{Conclusion}
In this work we have studied the coupling of YSR states in short one-dimensional chains of Fe atoms on clean Nb(110). We showed that functionalizing our STM tips with CO molecules significantly increases the spatial resolution of our experiment, allowing us to identify single atoms in nearest-neighbor lattice positions. Spectroscopic measurements exposed splitting of the single-atom YSR states for all investigated chains. By spatially resolving the LDOS of these states in differential conductance maps we uncovered their spatial symmetry and revealed a reversal of the odd and even state's energy position as a function of chain length. We were able to rationalize these findings with a simple tight-binding model. We believe that our results provide a comprehensive understanding of the coupling of YSR states in multi-impurity systems and constitute a step toward engineering more complex YSR systems~\cite{rontynen2015,korber2018}.

\begin{acknowledgments}
We would like to thank D. K. Morr for fruitful discussions.
The work was supported by the DFG through SFB 1180 (project C02).  
We acknowledge financial support by the Deutsche Forschungsgemeinschaft (DFG, German Research Foundation) 
under Germany's Excellence Strategy through W{\"u}rzburg-Dresden Cluster of Excellence 
on Complexity and Topology in Quantum Matter -- ct.qmat (EXC 2147, project-id 390858490).
\end{acknowledgments}

\bibliography{References_v4}

\begin{thebibliography}{43}
\bibitem [{ato()}]{atomManipulation}%
  \BibitemOpen
  \href@noop {} {}\bibinfo {note} {The attempt to use atomic manipulation of Fe
  atoms to create a \hkl[001]-oriented dimer was unsuccessful. Since we cannot
  pick up CO molecules from the Nb surface, we could not resolve the created
  dimers, however spectroscopic measurements on the manipulated dimers showed
  features very similar to the \hkl[1-11]-oriented dimer. We hence assume that
  an attractive force between the Fe atoms prevents the formation of
  \hkl[001]-oriented dimers as the atoms are pulled into nearest-neighbor
  lattice sites.}\BibitemShut {Stop}%
\bibitem [{oxy()}]{oxygenInfluence}%
  \BibitemOpen
  \href@noop {} {}\bibinfo {note} {Only for Fe atoms sitting on top of oxygen
  reconstructed areas the YSR states are altered, always accompanied by a
  change of the apparent height of the atoms. Already at distances smaller than
  $2\cdot a$, no influence of oxygen on the superconducting LDOS of the Fe
  atoms is observable.}\BibitemShut {Stop}%
\bibitem [{hop()}]{hoppingTerm}%
  \BibitemOpen
  \href@noop {} {}\bibinfo {note} {The value for $k_1$ is obtained by averaging
  over the peak positions at positive bias for several \hkl[1-11]-oriented
  dimers. Due to the weak intensity of the odd state, the extracted energy of
  this state varies for different dimers and is not always equal for positive
  and negative bias. Including the states at negative bias slightly shifts the
  average value of $k_1$ to
  $(0.43\pm0.05)\,\mathrm{meV}\approx(0.28\pm0.03)\,\Delta_\mathrm{Nb}$. This
  does not diminish the agreement of our calculations with the presented
  data.}\BibitemShut {Stop}%
\end{thebibliography}

\end{document}


\title{Supplementary Material for ``Coupling of Yu-Shiba-Rusinov states in 1D chains of Fe atoms on Nb(110)"}
\author{Felix Friedrich}
	\email[corresponding author: \\]{felix.friedrich@physik.uni-wuerzburg.de}
	\affiliation{Physikalisches Institut, Experimentelle Physik II,
Universit\"at W\"urzburg, Am Hubland, 97074 W\"urzburg, Germany}
\author{Robin Boshuis}
\affiliation{Physikalisches Institut, Experimentelle Physik II,
Universit\"at W\"urzburg, Am Hubland, 97074 W\"urzburg, Germany}
\author{Matthias Bode}
\affiliation{Physikalisches Institut, Experimentelle Physik II,
Universit\"at W\"urzburg, Am Hubland, 97074 W\"urzburg, Germany}
\affiliation{Wilhelm Conrad R\"ontgen-Center for Complex Material Systems (RCCM), Universit\"at W\"urzburg, Am Hubland, 97074 W\"urzburg, Germany}
\author{Artem Odobesko}
\affiliation{Physikalisches Institut, Experimentelle Physik II,
Universit\"at W\"urzburg, Am Hubland, 97074 W\"urzburg, Germany}

\date{\today}

\pacs{}
\maketitle

\vfill\pagebreak
{\noindent\bf{Supplementary Note I. Supplementary differential conductance maps}} \\
\begin{figure}[h]
\includegraphics[scale=1]{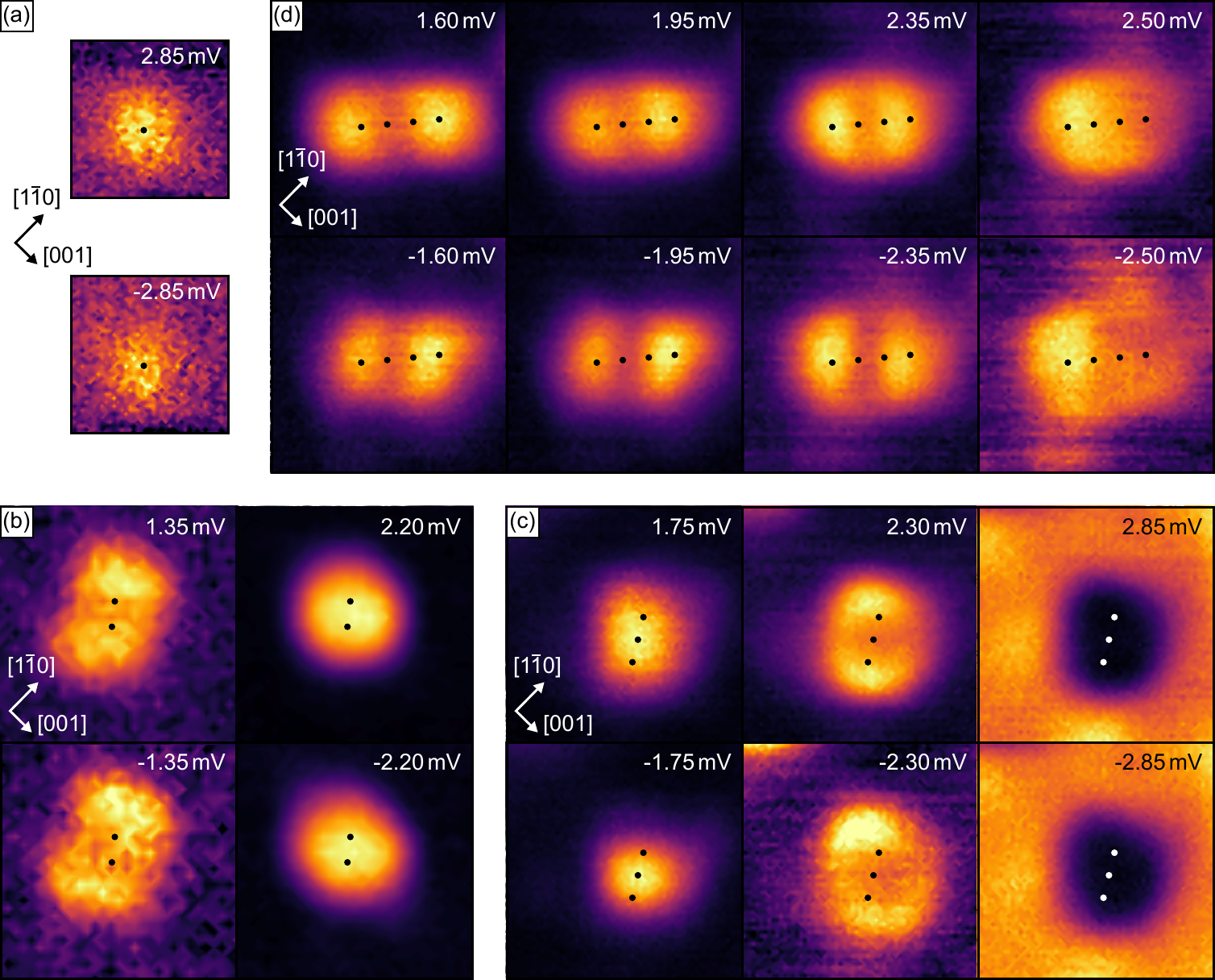}
\caption{Differential conductance maps for the electron- (upper panels) and hole-like states (lower panels) in a (a) single Fe atom (scan range: $2\times2\,\mathrm{nm^2}$), (b) dimer, (c) trimer and (d) tetramer (scan range: $3\times3\,\mathrm{nm^2}$). Since the YSR state of the single Fe atom only leads to a small enhancement of the coherence peak, the increase of the $\dd I/\dd U$ signal around the atom is very weak. For the trimer, in addition to the maps shown in Fig.\ 3 of the main text, we show a map at the bias voltage corresponding to the energetic position of the coherence peaks $(U=\pm2.85\,\mathrm{mV})$, where we expect the third YSR state. As discussed in the main text, the intensity of the coherence peaks on top of the chain is reduced in comparison with the clean surface due to the presence of in-gap states. The supplementary maps for the tetramer $(U=\pm1.95\,\mathrm{mV}, \pm2.35\,\mathrm{mV})$ in (d) correspond to the possible (spectrally unresolved) peaks discussed in the main text. We find that both show a vanishing intensity at the chain center, but attribute this to the much more intense state at $\pm2.35\,\mathrm{mV}$. Therefore, the symmetry of the state at $\pm1.95\,\mathrm{mV}$ remains unresolved. Note that the maps on the dimer were recorded with a different tip ($\Delta_\mathrm{tip}=1.31\,\mathrm{meV}$) than the other maps shown here ($\Delta_\mathrm{tip}=1.33\,\mathrm{meV}$). The maps in (c) and (d) are the same as those presented in Fig.\ 3 of the main text.}
\label{fig:didumaps}
\end{figure} \\

\vfill\pagebreak

{\noindent\bf{Supplementary Note II. Constant-current topography of Fe chains}} \\
\begin{figure}[h]
\includegraphics[scale=1]{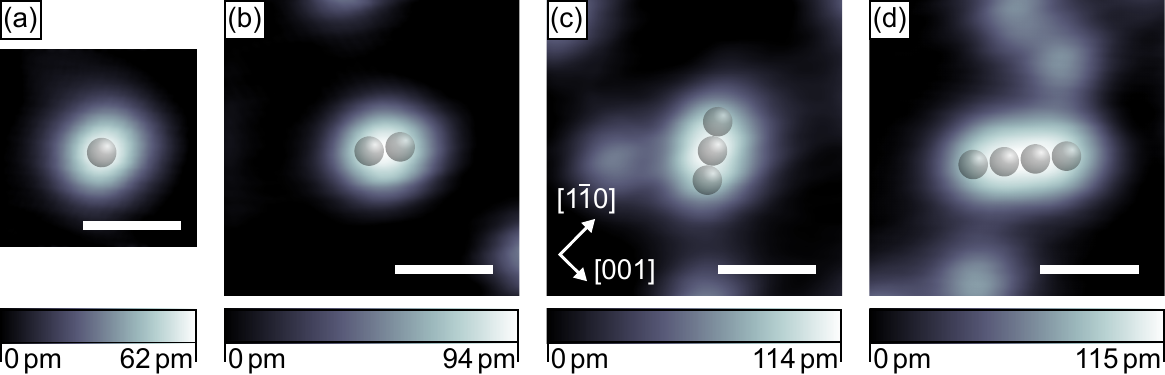}
\caption{Constant-current topographic images corresponding to the $\dd I/\dd U$ maps of (a) the single Fe atom in Fig.~\ref{fig:didumaps}(a) and (b)-(d) the Fe chains in Fig.\ 3 of the main text measured with a superconducting tip (scale bar is $1\,\mathrm{nm}$). Scan parameters: $U=-7\,\mathrm{mV}, I=1\,\mathrm{nA}$.}
\label{fig:COspectra}
\end{figure} \\

\vfill\pagebreak

{\noindent\bf{Supplementary Note III. Fe atom adsorption positions}} \\
\begin{figure}[h]
\includegraphics[scale=1.15]{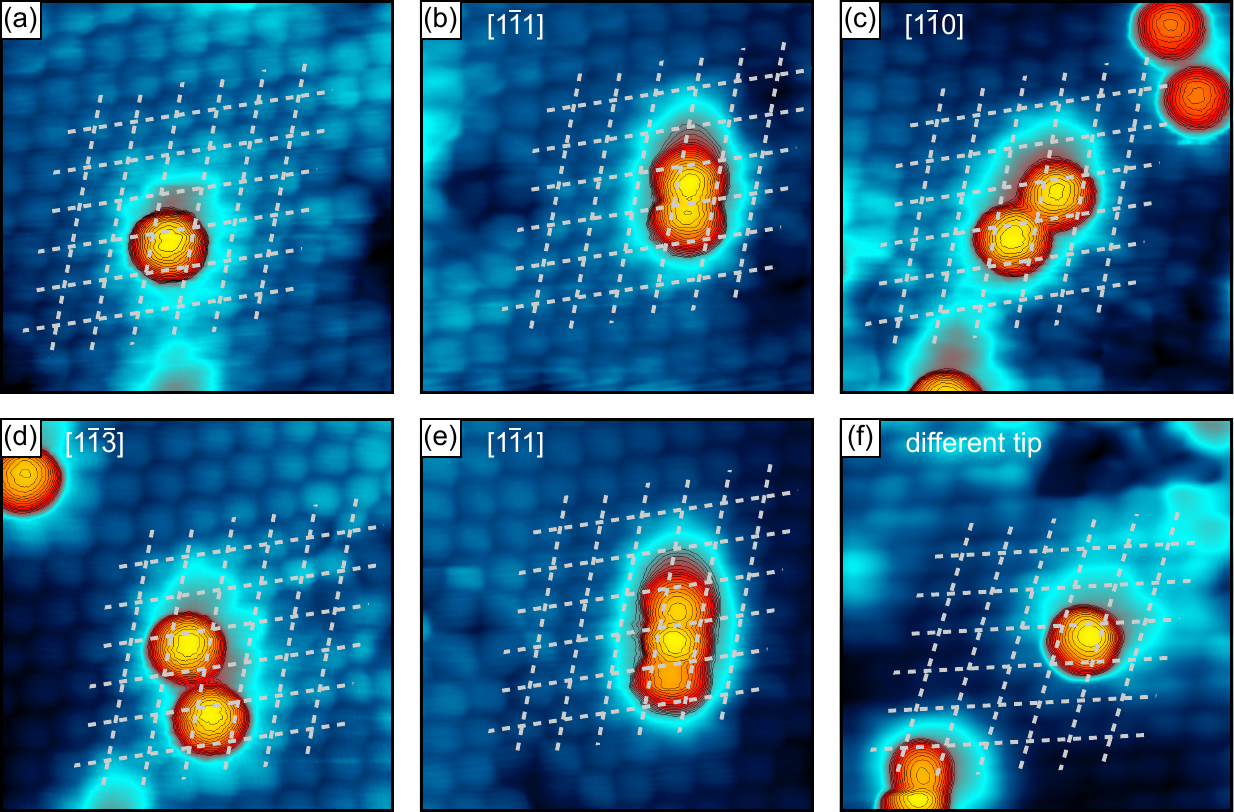}
\caption{Constant-current topographic images of (a),(f) single Fe atoms, (b)-(d) various dimers and (e) a nearest-neighbor trimer measured with CO-functionalized tips (scan range $3\times3\,\mathrm{nm^2}$). The data presented in panels (a)-(e) were recorded with the same tip, a different tip was used for (f). The  orientation of the dimers and trimer are indicated in the top left corner of every scan. The grey dashed lines follow the Nb(110) lattice. Depending on the tip, the atoms seem to be shifted from the four-fold hollow site [cf.\ (a),(f)]. With several different tips, no preferred shift direction is observed, while for the same tip, the shift is always the same. We hence follow that the single atoms are actually sitting in the low energy four-fold hollow site predicted by density functional theory calculations~\cite{odobesko2020}. For the nearest-neighbor chains described in the main text we additionally observe a shift of the atoms' LDOS toward the chain center [cf.\ (b),(e)], independent of the tip. Atoms in clusters with larger inter-atomic distance do not show this shift [cf.\ (c),(d)]. Scan parameters: $U=-10\,\mathrm{mV}, I=5\,\mathrm{nA}$.}
\label{fig:NbOx_phaseI}
\end{figure} \\

\vfill\pagebreak

{\noindent\bf{Supplementary Note IV. Inelastic tunneling spectroscopy of CO-functionlized tips}} \\
\begin{figure}[h]
\includegraphics[scale=1]{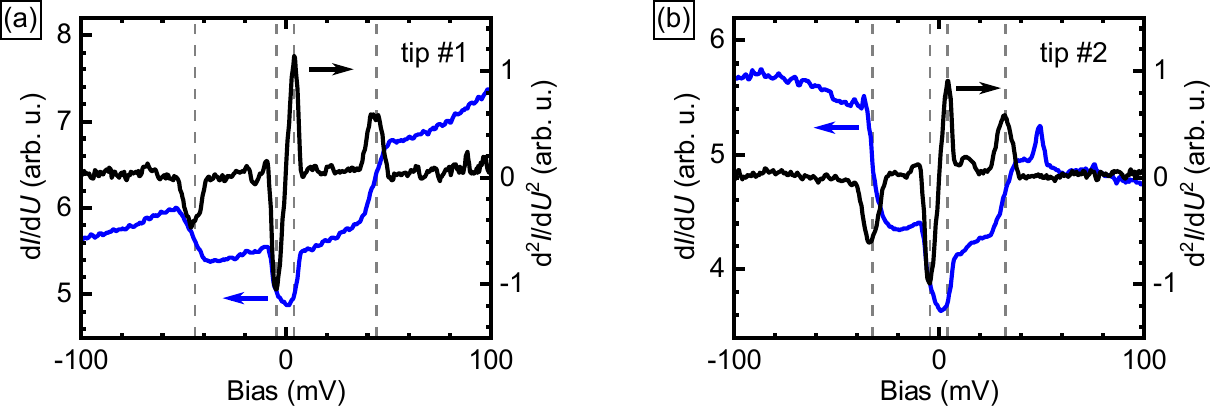}
\caption{Measured first (blue curve) and second (black curve) derivative of the tunneling current of two different CO-functionalized tips on top of clean Nb(110). Dashed lines mark the excitation energy of the hindered translation (low energy) and hindered rotation mode (high energy) of the CO molecule on the STM tip. As can be seen from the difference of (a) and (b), this energy is tip/adsorption position-dependent. We obtain (a) $4\pm1\,\mathrm{meV}$ and $44\pm2\,\mathrm{meV}$ and (b) $4\pm1\,\mathrm{meV}$ and $33\pm2\,\mathrm{meV}$ for both modes, respectively. Stabilization parameters: $U=-100\,\mathrm{mV}$, $I=1\,\mathrm{nA}$, $U_\mathrm{mod}=5\,\mathrm{mV}$.}
\label{fig:COspectra}
\end{figure} \\

\vfill\pagebreak

{\noindent\bf{Supplementary Note V. Modeling the coupling of YSR states}}

In this section we briefly review the simplified tight-binding model used for the results in the main text and further discuss more rigorous calculations introduced in Ref.~\onlinecite{ruby2018}. Although these calculations are expected to yield more accurate results, they also require several input parameters such as the Fermi wavelength, the strength of the adatom--substrate interaction and the YSR wave function that are unavailable for our system.

Following the description in Ref.~\onlinecite{ruby2018}, the Hamiltonian of the $n$-atom system is given by
\begin{align}
H = H_\mathrm{s}+\sum_i^nJ_\mathrm{i}(r),\label{eq:hamilton}
\end{align}
where $H_\mathrm{s}$ describes the superconductor and $J_\mathrm{i}(r):=J(r-r_\mathrm{i})$ reflects the exchange potential between an adatom at site $r_\mathrm{i}$ and the itinerant quasiparticles. We here omit the spin-dependent part of the Hamiltonian~\cite{ruby2018}. We assume that the system for $n=1$ is solved by the single-atom YSR wave function $\varphi_\mathrm{i}:=\varphi(r-r_\mathrm{i})$ and the single-atom YSR energy~$E_0$. For small interactions the solution is based on a linear combination the single-atom solutions:
\begin{align}
\Phi(r)=\sum_i^n c_\mathrm{i}\varphi_\mathrm{i}.\label{eq:ansatz}
\end{align}
The solution of the two-impurity system
\begin{align}
\begin{pmatrix}
E_0+C & E_0S+D\\
E_0S+D & E_0+C
\end{pmatrix}
\begin{pmatrix}
c_1\\
c_2
\end{pmatrix}
=E
\begin{pmatrix}
1 & S\\
S & 1
\end{pmatrix}
\begin{pmatrix}
c_1\\
c_2
\end{pmatrix}
\end{align}
is given by
\begin{align}
E=E_0+\frac{C\pm D}{1\pm S},
\end{align}
with the overlap integral~$S$, the Coulomb-like integral~$C$, and the exchange-like integral~$D$ as given in Ref.~\onlinecite{ruby2018}. As mentioned above, the calculation of these integrals requires knowledge of the exact wave functions, including the Fermi wavelength, and the exchange potential. However, we can neglect $S$ if the overlap between the wave functions is small. Further, as $C$ rigidly offsets both solutions equally and we are interested in the energetic splitting of the YSR states only, we can further omit $C$. Finally, by replacing the integral~$D$ with the hopping term~$k_1$ used in the main text, the problem reduces to
\begin{align}
\begin{pmatrix}
E_0 & k_1\\
k_1 & E_0
\end{pmatrix}
\begin{pmatrix}
c_1\\
c_2
\end{pmatrix}
=E
\begin{pmatrix}
c_1\\
c_2
\end{pmatrix}
\end{align}
with solutions 
\begin{align}
E_\mathrm{dimer,e}=E_0+k_1,\\
E_\mathrm{dimer,o}=E_0-k_1,
\end{align}
as already shown in Ref.~\onlinecite{morr2003,morr2006}. The indices ``e'' and ``o'' indicate the even or odd spatial symmetry of the solutions, respectively. By introducing the next-nearest and next-next-nearest neighbor hopping terms $k_2$ and $k_3$ for the trimer and tetramer we obtain:
\begin{align}
E_\mathrm{trimer,e}&=E_0+\frac{k_2}{2}\pm\sqrt{2k_1^2+\frac{k_2^2}{4}},\\
E_\mathrm{trimer,o}&=E_0- k_2,\\
E_\mathrm{tetramer,e}&=E_0+\frac{k_1}{2}+\frac{k_3}{2}\pm\sqrt{\frac{5k_1^2}{4}+2k_1k_2-\frac{k_1k_3}{2}+k_2^2+k_3^3},\\
E_\mathrm{tetramer,o}&=E_0-\frac{k_1}{2}-\frac{k_3}{2}\pm\sqrt{\frac{5k_1^2}{4}-2k_1k_2-\frac{k_1k_3}{2}+k_2^2+k_3^3}.
\end{align}
As discussed in the main text we can neglect $k_2$ and $k_3$ for our system. The results of the remaining equations are displayed in Fig.\ 4 of the main text.

\begin{figure}
\includegraphics[scale=1]{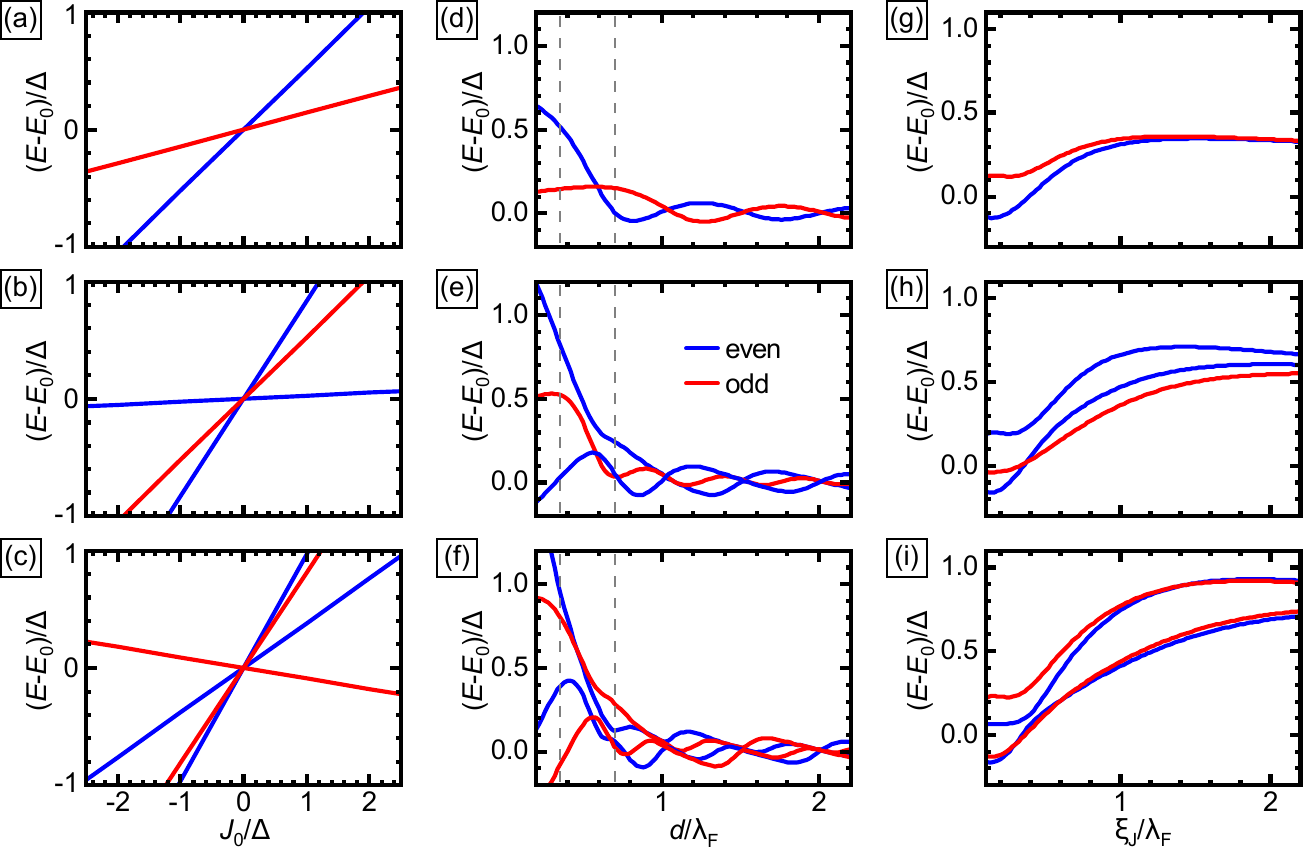}
\caption{Calculated energy shifts from the single-atom energy~$E_0$ depending on (a)-(c) the exchange potential strength, (d)-(f) the inter-atomic distance and (g)-(i) the decay length of the exchange potential for a dimer, trimer and tetramer (top to bottom). See text for details on the model. The symmetry of the states (even/odd) is reflected by the line color (blue/red). Parameters used: (a)-(c) $\xi_\mathrm{J}/\lambda_\mathrm{F}=0.4$, $d/\lambda_\mathrm{F}=0.35$; (d)-(f) $\xi_\mathrm{J}/\lambda_\mathrm{F}=0.4$, $J_0/\Delta=1$; (g)-(i) $d/\lambda_\mathrm{F}=0.7$, $J_0/\Delta=1$. The dashed lines in (d)-(f) indicate the inter-atomic distances used in (a)-(c) and (g)-(i), respectively.}
\label{fig:frankeModel}
\end{figure}

To assess the validity of our simplified calculations we exemplarily solve the Schr{\"o}dinger equation with the Hamiltonian in Eq.~\eqref{eq:hamilton} and the ansatz in \eqref{eq:ansatz} with an exchange potential of the form~\cite{ruby2018} \mbox{$J(r)=J_0/\sqrt{\pi\xi_\mathrm{J}}\exp(-r^2/\xi_\mathrm{J}^2)$} and a normalized YSR wave function of the form~\cite{rusinov1969} \mbox{$\sqrt{2/\lambda_\mathrm{F}}\mathrm{\ sinc}(2\pi r/\lambda_\mathrm{F})$} in one dimension, where $\lambda_\mathrm{F}$ is the Fermi wavelength and $r$ points along the chain. We display the resulting YSR energies for the dimer, trimer and tetramer as a function of the exchange potential strength~$J_0/\Delta$, the inter-atomic distance~$d/\lambda_\mathrm{F}$ and the exchange potential decay length~$\xi_\mathrm{J}/\lambda_\mathrm{F}$ in Fig.~\ref{fig:frankeModel}. Similar to the reduced model, the YSR energies increase linearly with the exchange potential strength. However, the splitting is not symmetric around the single-atom YSR energy, but is accompanied by a positive shift of the solutions due to the overlap of the wave functions and the Coulomb-like terms in the solution. Further, the energy difference between energetically neighboring states varies within one chain configuration [cf.\ panel (b) or (c)]. Yet, the most striking difference between the simplified and the extended calculation is reflected in the distance dependence shown in Fig.~\ref{fig:frankeModel}(d)-(f). Whereas for small distances even and odd states alternate as a function of energy, this is not true for all distances. The distance-dependent oscillation of the YSR state energies, which is a consequence of the phase difference between the single-atom wave functions, reverses or completely mixes the order of odd and even states. For example, at an inter-atomic distance of $0.7\lambda_\mathrm{F}$ [second dashed line in panels (d)-(f)] the two energetically highest states are both even for the trimer, while for the tetramer the lowest and highest state are odd. In addition, the decay length of the exchange potential [panel (g)-(i)] can also change the energetic order of the even and odd states.

From the results discussed here we can deduce the limits of applicability of the simplified model to real systems. As revealed by the extended calculations, the alternation between odd and even states as a function of energy, that is also found in the simplified model, only occurs for small inter-atomic distances, where the phase difference between the wave functions is negligible, or for sufficiently large inter-atomic distances, where the phase difference is well defined. In addition, the simplified model is only valid if the exchange potential is localized to small distances around the magnetic impurity, as indicated by the limited range of $\xi_\mathrm{J}$ over which the energetic order of the states is maintained [cf.\ Fig.~\ref{fig:frankeModel}(g)-(i)]. We assume that our system  of Fe atoms on Nb(110) is in the limit of inter-atomic distances much smaller than $\lambda_\mathrm{F}$. Our measurements further suggest a very limited range of inter-atomic exchange since the coupling of YSR states is restrained to nearest-neighbor atoms.

A quantitative comparison between both calculations shows that the simplified model can only yield approximations of the YSR energies. Although in both models the YSR energies change linearly with the exchange potential strength, the relative energy differences between the states depend on the exact ratios of $d/\lambda_\mathrm{F}$ and $\xi_\mathrm{J}/\lambda_\mathrm{F}$ as well as the form of the single-atom YSR wave function in the extended calculations. We want to emphasize that the extended model also does not provide the full energy shift, even if all necessary input parameters are known; among others the spin alignment of the magnetic atoms in the chain is neglected. For small inter-atomic coupling as observed in our experiments, the simplified model yields satisfactory agreement with the experimental data without the need for further modeling of the system to obtain e.g. the atom--substrate interaction or \mbox{the surface band structure.}

\bibliography{References_v3}